# Investigating the spatial heterogeneity of factors influencing speeding-related crash severities using correlated random parameter order models with heterogeneity-in-means

**Renteng Yuan (First author)**
Jiangsu Key Laboratory of Urban ITS
School of Transportation
Southeast University, Nanjing, Jiangsu, P. R. China, and 210000
Email: rtengyuan123@126.com

**Qiaojun Xiang (Corresponding author)**
Jiangsu Key Laboratory of Urban ITS
School of Transportation
Southeast University, Nanjing, Jiangsu, P. R. China, and 210000
Email: 230208815@seu.edu.cn

**Zhiheng Fang**
Jiangsu Key Laboratory of Urban ITS
School of Transportation
Southeast University, Nanjing, Jiangsu, P. R. China, and 210000
Email: 220213476@seu.edu.cn

**Xin Gu**
Beijing Key Laboratory of Traffic Engineering
Beijing University of Technology, Beijing, P. R. China, and 100124
Email: guxin@bjut.edu.cn

**Abstract:**
Speeding has been acknowledged as a critical determinant in increasing the risk of crashes and their resulting injury severities. This paper employs Global Moran's I coefficient and local Getis–Ord G* indexes to systematically account for the spatial distribution feature of speeding-related crashes, study the global spatial pattern of speeding-related crashes, and identify severe crash cluster districts. The findings demonstrate that severe speeding-related crashes within the state of Pennsylvania have a spatial clustering trend, where four crash datasets are extracted from four hotspot districts. Two log-likelihood ratio (LR) tests were conducted to determine whether speeding-related crashes classified by hotspot districts should be modeled separately. The results suggest that separate modeling is necessary. To capture the unobserved heterogeneity, four correlated random parameter order models with heterogeneity in means are employed to explore the factors contributing to crash severity involving at least one vehicle speeding. Overall, the findings exhibit that some indicators are observed to be spatial instability, including hit pedestrian crashes, head-on crashes, speed limits, work zones, light conditions (dark), rural areas, older drivers, running stop signs, and running red lights. Moreover, drunk driving, exceeding the speed limit, and being unbelted present relative spatial stability in four district models. This paper provides insights into preventing speeding-related crashes and potentially facilitating the development of corresponding crash injury mitigation policies.

**Keyword:** Crash severity, Speeding-related crash, Correlated random parameter model, Heterogeneity in means

# 1. Introduction

Speeding behavior has often been regarded as the main primary to crashes and has long been a concern for safety researchers (Aarts and van Schagen 2006; Ahie et al. 2015; Mohamad et al. 2019; Shyhalla 2014). In 2020, there were 38,824 fatalities on U.S. roadways, of which, 11,258 (29%) involved at least one vehicle speeding, where speeding-related fatalities accounted for 40.7% of the total(NHTSA 2020). Speeding-related fatalities accounted for 41% (459) of the total 1,129 fatalities in the state of Pennsylvania in 2020 (NHTSA,2020). Therefore, there is an urgent need to develop effective safety strategies to reduce the crash occurrence and minimize crash consequences.

Research has shown that vehicle speed varies in different road segments (Aarts and van Schagen 2006; Ahie et al. 2015; Wang et al. 2015). Road segments with lower speed limits, central business district areas, secondary roads, lower traffic volume were positively correlated with high rates of speeding(Huang et al. 2018)(Wang et al., 2018). Speeding behavior increases the risk of crashes and affects crash severity(Aarts and van Schagen 2006).Trends in the clustering of speeding driving behavior may impact the spatial distribution characteristics of speeding-related crashes. Previous studies have confirmed that such clustering area spatial heterogeneity may result in significant variability in the factors affecting crash severity(Song et al. 2020; Wang et al. 2022b; Yan et al. 2021b). Exploring the spatial distribution pattern of speeding-related crashes is crucial for analyzing crash severity.

A considerable amount of research has been conducted on the macro (zonal) level (e.g., the developing country, the county level, the city level, mountainous areas, rural areas, and work zones) to meet the needs of region-level traffic safety management(Huang et al. 2008; Lee and Li 2014; Li et al. 2013; Rahman et al. 2021; Scheiner and Holz-Rau 2011; Se et al. 2021a; Zeng et al. 2022). Recently, traffic safety analyses at the micro level (e.g., segments, and intersections) have received much attention(Adanu et al. 2021; Afghari et al. 2020; Alarifi et al. 2018; Bhowmik et al. 2021; Bhowmik et al. 2018; Chen et al. 2016b; Guo et al. 2019; Rezapour and Ksaibati 2022; Wang et al. 2021). At the same time, the limitations of these studies are gradually being reported. For instance, many studies have reported that rural areas increase crash severity and are clusters of severe crashes (Islam and Burton 2019; Se et al. 2021a; Wu et al. 2014). However, some research has also found that not all rural roads are positively associated with an increased likelihood of severe crashes influenced by weather, light conditions, traffic volume, and traffic management measures(Alrejjal et al. 2022; Zhang and Hassan 2019b). Therefore, capturing spatial heterogeneity, breaking down the barriers for regional traffic safety analysis (e.g., regional constraints by administrative, urban, and rural areas), and identifying severe crash cluster districts facilitates the accurate prevention and control of traffic accidents. Since previous studies have shown that speeding driving behavior and the injury severity of crashes are, to some extent, induced by factors that change over space(Se et al. 2022; Song et al. 2020), this study is particularly interested in the following problems:

(1) What is the spatial distribution pattern of speeding-related crashes? How are speeding-related crash hotspot areas identified?
(2) What are the determinants of speeding-related crash severity? Do significant factors vary across different crash hotspot areas?

The rest of the paper is organized as follows. Section 2 presents a literature review, Section 3 outlines the detailed methodology design, Section 4 describes the dataset employed in the study, Section 5 conducts two spatial stability tests, Section 6 presents the research findings, and Section 7 summarizes the discussions and conclusions drawn from the study.

**2. Literature review**

An accident will be considered a speeding-related crash if any driver is records as driving too fast for the environment conditions, or exceeding the speed limit of the road (NHTSA 2020; Se et al. 2022). This section summarizes the literature from three aspects: research on speeding-related crash severity, spatial distribution identification methods, and unobserved heterogeneity.

**2.1 Related work on speeding-related crash severity**

Many studies have been conducted with "speeding driving" as an explanatory variable in the crash severity modeling process(Anarkooli et al. 2017; Rezapour et al. 2019; Se et al. 2021b). These studies concluded that speeding significantly affects crash severity but failed to identify the specific factors affecting the severity of speeding-related crashes. As exemplified by **(Lemp et al. 2011)**, an investigation into the effects of vehicle, occupant, driver, and environmental characteristics on injury outcomes among individuals involved in heavy-duty truck accidents revealed that speeding by any vehicle immediately prior to the crash is associated with heightened crash severity. Osman et al. (2018) centered their research on single-vehicle crashes involving drivers with commercial licenses and concluded that there is a strong positive correlation between speeding and the severity of such crashes. In recent years, very few studies have focused on investigating the factors affecting speeding-related crash severity(Hoye 2020; Islam and Mannering 2021; Se et al. 2022). These studies have been fruitful in identifying the significant factors determining speeding-related crash severity. For instance, the study conducted by Se et al. (2022) aimed to explore the dissimilarities in the temporal stability of factors that affect driver-injury severities in crashes involving speeding and non-speeding driving. Islam and Mannering (2021) investigate the role of gender and temporal instability in driver-injury severities in crashes caused by speeds too fast for conditions. However, no study has been conducted to comprehensively evaluate crash severity considering heterogeneity in the spatial distribution of speeding-related crashes. Ignoring such spatial heterogeneity in speeding-related crash severity analysis may lead to biased results. Therefore, there is a need for more extensive research on speeding-related crash severity.

**2.2 Related work on approaches for the spatial distribution of crashes**

Spatial autocorrelation analysis includes both global and local analyses. Before performing local spatial autocorrelation analysis, it is necessary to test the spatial distribution pattern of accidents (global analysis). Nearest neighbor distances, kernel density estimation, and K-function have been developed for the global spatial autocorrelation analysis(Yamada and Thill 2004), but are limited to equal weights for each point(Songchitruksa and Zeng 2010). Global Moran's I coefficient, proposed by Griffith (1987), incorporates both the location information and attribute value of points. It has been widely used in global spatial autocorrelation analyses to reflect the spatial distribution pattern of attribute variables(Xiao et al. 2018). For the local spatial autocorrelation analysis (hotspot analysis), the Getis-Ord G* index and the kernel density estimation (KDE) were employed to identify the "hot" areas of the spatial distribution(Alrejjal et al. 2022; Li and Fan 2020; Pulugurtha et al. 2007; Xiao et al. 2018). Previous studies have indicated that KDE is infeasible for the statistical test(Song et al. 2021a; Song et al. 2020). The Getis-Ord G* index has been widely used to identify locations with spatially aggregated crashes (Pulugurtha et al. 2007; Song et al. 2020; Song et al. 2021b)), which requires the aggregation of data based on the number of crashes in an area under study divided into smaller space units called rasters(Plug et al. 2011).

**2.3 Related work on approaches for unobserved heterogeneity**

In terms of methodology, it is important to note that crash severity is frequently recorded as an ordinal scale variable (e.g., no injury, minor injury, and serious injury). In analyzing crash severity data, two primary modeling methods have been employed: ordered(Eluru et al. 2008; Lemp et al. 2011; Qi et al. 2013; Yuan et al. 2022b) and unordered(Hou et al. 2019; Tay et al. 2011; Wang et al. 2022b; Yuan et al. 2022a). Regarding the ordered frameworks, the unordered model fails to account for the natural ordering of injury severity outcomes, which may lead to misleading or inaccurate results (Osman et al. 2018). Many factors significantly affect speeding-related crash severity, such as the driver, vehicle, road infrastructure, and traffic environment. It is unrealistic to consider all factors when modeling crash severity, which can lead to unobserved heterogeneity(Behnood and Mannering 2019; Mannering et al. 2016; Wang et al. 2022a).The random parameter modeling approach which includes the random parameter ordered probit (RP-Probit) model, the RP-Probit model with heterogeneity in means and variance (RPMV-Probit), and the correlated RP-Probit model with heterogeneity in means (CRPM-Probit), has been developed to capture unobserved heterogeneity. Previous studies compare the statistical performance of uncorrelated and correlated random-parameter models, reporting that the correlated random-parameter model is statistically superior to uncorrelated models(Saeed et al. 2019; Se et al. 2021a; Se et al. 2021b). Therefore, the CRPM-Probit models with heterogeneity in means are conducted in this paper to capture unobserved heterogeneity and provide insights into crash prevention countermeasures.

**3. Methodology**
**3.1 Global spatial autocorrelation analysis**
Global spatial autocorrelation analysis is mainly focused on identifying the global spatial pattern of space units and investigating their trends. The global spatial pattern contains clustering patterns, discrete patterns and random patterns(Songchitruksa and Zeng 2010). Local spatial autocorrelation analysis is not required if crashes are randomly distributed in global analysis(Le et al. 2020). In this paper, Global Moran's I coefficient is selected to determine the spatial distribution of speeding-related crashes comes. Global Moran's I is expressed as:

$$I = \frac{\sum_{i=1}^{n}\sum_{j=1}^{n} \omega_{ij}(x_i - \bar{x})(x_j - \bar{x})}{\left(\sum_{i=1}^{n}\sum_{j=1}^{n}\omega_{ij}\right)\sum_{i=1}^{n}(x_i - \bar{x})^2} \qquad (1)$$

Where $x_i$ is the attribute value in the i-th raster (space unit); $x_j$ is the attribute value in the j-th raster; $\bar{x}$ represents the average attribute value in the global space; $n$ denotes the number of space units of global region, $\omega_{ij}$ is the spatial weight matrix ($\omega_{ij} = 1$ if the j-th raster is in contiguity with the i-th raster (contiguity edges corners), and 0 otherwise). Global Moran's I coefficient belongs to [-1,1], and it will be positive if the pattern of spatial distribution is a clustering pattern (serious crashes tend to cluster in certain districts). Initially, Global Moran's I assumed that the elements in the global space are randomly distributed, and the Z-score of Moran's I is expressed as:

$$Z = \frac{I - E(I)}{\sqrt{V(I)}} \qquad (2)$$

Where V(I) is the variance of Moran's I, E(I) represents the expectation values of Moran's I and can be calculated as E[I]=-1/(n-1). The null hypothesis can be rejected when $|z| > 1.96, p < 0.05$, (Songchitruksa and Zeng 2010; Xiao et al. 2018).

**3.2 Hotspot Analysis**

Hotspot analysis is mainly focused on identifying the severe crash cluster districts, considering the spatial correlation between a raster and its neighboring units. In this paper, the local Getis-Ord G* index is selected for local spatial autocorrelation analysis as in previous studies (Song et al. 2020; Songchitruksa and Zeng 2010). The local Getis–Ord G* index is defined as:

$$G_i^* = \frac{\sum_{j=1}^{n}\omega_{ij}x_j - \bar{x}\sum_{j=1}^{n}\omega_{ij}}{\sqrt{\frac{\sum_{j=1}^{n}x_j^2}{n} - \bar{x}^2} \times \sqrt{\left[n\sum_{j=1}^{n}\omega_{ij}^2 - \left(\sum_{j=1}^{n}\omega_{ij}\right)^2\right]/(n-1)}} \qquad (3)$$

Where the local Getis–Ord G* index is a statistic Z-score, all parameters are explained as described above. For statistically significant positive p-values, a higher score (G*>0) indicates a tighter clustering of high values (hot spots), a lower G* score (G*<0) denotes a tighter clustering of lower values (cold spots). A statistically insignificant p-value or G*=0 means a random pattern of the values.

**3.3 Correlated random parameter order model with heterogeneity in means**

In this research, the CRP-Probit model with heterogeneity in means is employed. At first, the traditional ordered probit model is defined as follow:

$$y_i^* = \beta_i x_i + \varepsilon_i \qquad (4)$$

Where $x_i$ represents a vector of explanatory variables for the i-level crash severity, $\beta_i$ represents a vector of estimable parameters, $\varepsilon_i$ is random error term and assumed to be normally distributed, $y_i^*$ is a latent continuous variable that is defined as:

$$y_i = j, \text{ if } u_{j-1} < y_i^* < u_j \tag{5}$$

Where $y_i$ is the observe injury severity, $u_j$ represent the threshold parameters, j is the integers representing the injury-severity levels (j = 0, 1, 2, respectively, for no injury, minor, and severe injuries). The probability that crash i being j-th injury-severity, $p(y = j)$, is define as:

$$\begin{aligned} P(y=0) &= \phi(-\beta_i x_i) \\ P(y=1) &= \phi(u_1 - \beta_i x_i) - \phi(-\beta_i x_i) \\ P(y=2) &= 1 - \phi(u_1 - \beta_i x_i) \end{aligned} \tag{6}$$

Where $\phi$ is the cumulative standard normal distribution. A RP-Probit model is developed taking into account the effect of unobserved factors. Partial parameters $\beta_i$ are relaxed and allowed to vary across individual crash observations, as follows:

$$\beta_i = \beta + \zeta_i \tag{7}$$

Where $\beta$ is the constant term representing the mean value of the random parameters vector, $\zeta_i$ is the randomly distributed term (e.g. a normally distributed term with mean=0 and variance $\sigma^2$). Eq. (7) can be extended further by considering interaction effects, as follows:

$$\beta_i = \beta + \eta z_i + \Gamma \omega_i \tag{8}$$

Where $z_i$ is the vector of independent variables corresponding to random parameters $\beta_i$, $\eta$ represents a vector of estimable parameters, $\eta z_i$ is the heterogeneous term to capture unobserved heterogeneity resulting from interactions between random parameters and fixed parameters (Se et al. 2021b), $\omega_i$ represents a random term with a mean value of zero, $\Gamma$ is the Cholesky matrix(variance-covariance matrix). The random parameters are assumed to be mutually independent in the RP-Probit model, and the $\Gamma$ is defined as a diagonal matrix (Saeed et al. 2019), defined in Eq.(9):

$$\Gamma = \begin{bmatrix} \sigma_{1,1} & \sigma_{1,2} & \cdots & \sigma_{1,n-1} & \sigma_{1,n} \\ \sigma_{2,1} & \sigma_{2,2} & \cdots & \sigma_{2,n-1} & \sigma_{2,n} \\ \vdots & \vdots & \ddots & \vdots & \vdots \\ \sigma_{n-1,1} & \sigma_{n-1,2} & \cdots & \sigma_{n-1,n-1} & \sigma_{n-1,n} \\ \sigma_{n,1} & \sigma_{n,2} & \cdots & \sigma_{n,n-1} & \sigma_{n,n} \end{bmatrix} \tag{9}$$

Where $n$ represents the number of random parameters, $\sigma_{k,j}$ ($1 \leq k \leq n, 1 \leq j \leq n$) denotes the elements of the Cholesky matrix. The correlation between random parameters is considered in the CRPM-Probit model, and a generalized Cholesky matrix is defined in Eq.(10),.

$$Var(\beta_i | \omega_i) = \Gamma \Gamma' \tag{10}$$

The diagonal and off-diagonal elements of the $\Gamma$ matrix are not equal to zero in the generalized Cholesky matrix, and the standard deviations of the correlated random parameters can be computed based on Eq.(11).

$$\sigma_r = \sqrt{\sigma_{k,k}^2 + \sigma_{k,k-1}^2 + \sigma_{k,k-2}^2 + \cdots + \sigma_{k,1}^2} \tag{11}$$

Where $\sigma_r$ represents the standard deviation of the random parameter $r$, and $\sigma_{k,k}$ is the Cholesky matrix's diagonal element. The correlation coefficient between two random parameters can be determined as:

$$Cor(x_1,x_2) = \frac{Cov(x_1,x_2)}{\sigma_{x_1} \times \sigma_{x_2}} \tag{12}$$

Where $Cor(x_1,x_2)$ represents the correlation coefficient between $x_1$ and $x_2$, $Cov(x_1,x_2)$ denotes the covariance, $\sigma_{x_1}$ and $\sigma_{x_2}$ are the standard deviations.

In this paper, the estimated models include only binary indicators as independent variables, therefore, the change from "0" to "1" in the value of the variables determines the marginal effects(Fountas et al. 2021; Jalayer et al. 2018; Sadri et al. 2013; Se et al. 2021a), as:

$$\frac{\Delta P_i(y=j)}{\Delta x} = \left[\phi(\mu_{j-1} - \beta_i x_i) - \phi(\mu_j - \beta_i x_i)\right]\beta \tag{13}$$

Where $\phi$ represents the density function of the normal distribution, and all other terms remain defined earlier.

## 4 Data

The data used in this study were reported by law enforcement agencies from 2018 to 2020 and collected from the Pennsylvania Department of Transportation (PennDOT) Open Data Portal (https://crashinfo.penndot.gov/PCIT/welcome.html). For this study, only speeding-related crash records are extracted. Crash identification numbers are employed to link different data files such as crash, vehicle, person, and roadway characteristics. In the connection dataset, one crash record was considered a sample. Ultimately, a total number of 63736 crash records involving speeding was selected. In the original dataset, the injury severity was coded on the KABCO scale where K-level represented fatal injury, A-level represented suspected serious injury, B-level represented suspected minor injury, C-level represented possible injury, and O-level represented no apparent injury. In this paper, the maximum crash injury level is extracted. The dataset also provides detailed information about the crash location, including latitude, longitude, county, and regional characteristics (urban or rural. The spatial distribution of speeding-related crash data is illustrated in Fig. 1.

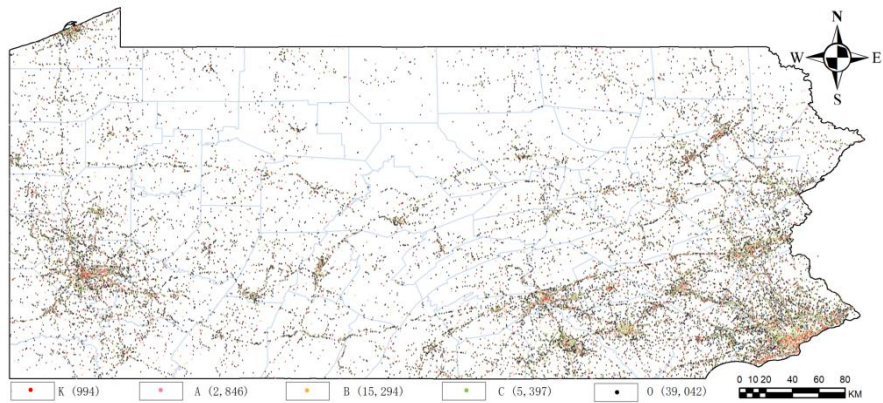

**Fig.1 Spatial distribution of speeding-related crashes**

Note: ( ) is the number of crashes

### 4.1 Data processing

The state of Pennsylvania is roughly rectangular in shape and stretches about 480 km from east to west and 240 km from north to south. It is further divided into 67,840 rasters (each raster area is about 1.86 km$^2$) to identify the global spatial pattern of space units. Then, the speeding-related crash data was converted into raster data. Rasters with a speeding-related crash count greater than 0(17,448 rasters) are retained in the dataset, while rasters with crash counts equal to 0 are removed. To accurately describe the casualty level of speeding-related crashes, the equivalent fatalities corresponding to the maximum injury severity are recommended to capture the equivalent impact of an injury relative to a fatality (K-level, A-level, B-level, C-level, and O-level are equivalent to 1, 0.1107, 0.0310, 0.0148, and 0.0049 fatalities, respectively) (Feng et al., 2016).

Global Moran's I coefficient is used to examine the spatial distribution pattern of speeding-related crashes. The corresponding Global Moran's Index is 0.012 (z-score = 9.17, P-value <0.001), indicating a clustering spatial pattern trend within the state of Pennsylvania. Accordingly, hotspot analysis is implemented to identify the district where high and low values of spatial units are clustered. Figure 2 illustrates the result of the Getis-Ord index in the state of Pennsylvania. The red areas represent hotspots, indicating that severe speeding-related crashes tend to cluster in those areas. The blue areas represent coldspots where no injury crashes tend to cluster. Four crash datasets are extracted from four hotspot districts to uncover the causal mechanisms of speeding-related crash severity in hotspot districts, as shown in Fig. 2. The maximum injury severity is aggregated into three injury severity categories: no injury (O-level), minor injury (B-level and C-level), and serious injury (K-level and A-level) in four extracted crash datasets (Islam, S. & Burton, 2019; Lee & Li, 2014). Table1 summarizes the descriptive statistic of the four crash datasets.

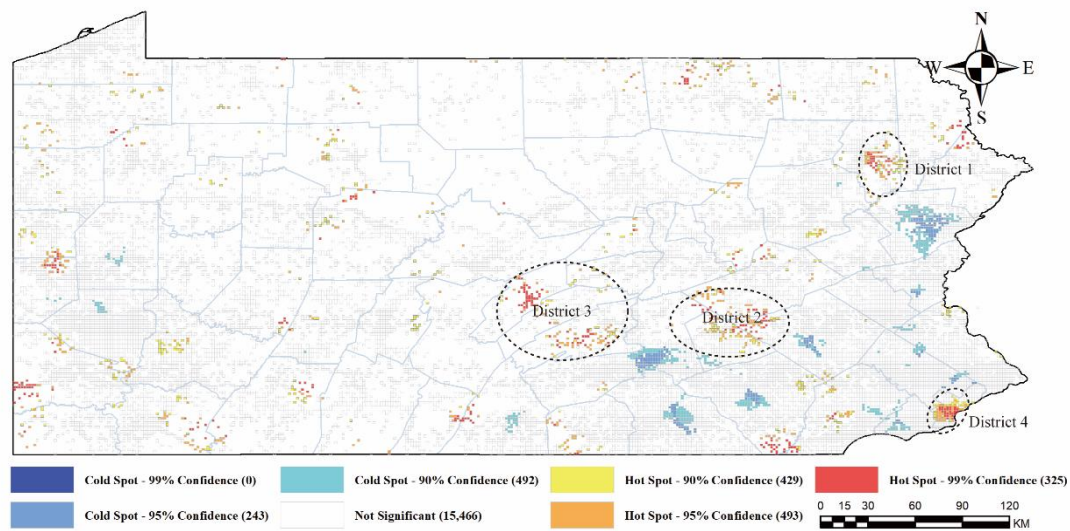

Fig.2. The results of Getis-Ord G* spatial statistics

Note: ( ) is the number of rasters

Fig. 3 presents the spatial distribution characteristics of the four districts, including land use, work zones, light condition, and intersection. It was observed that all crashes in District 4 and 73% of the crashes in District 1 occurred in urban areas, while 97% and 90% of the crashes occurred in rural areas in Districts 3 and 2, respectively. The analysis suggests that District 4 and 1 are primarily urban areas, whereas District 2 and 3 are predominantly rural areas. Similarly, Districts 1 and 2 have higher proportions of daylight crashes, while Districts 3 and 4 have a higher proportion of dark crashes. The analysis reveals that the road lighting conditions in Districts 3 and 4 are comparatively inferior to those in Districts 1 and 2.

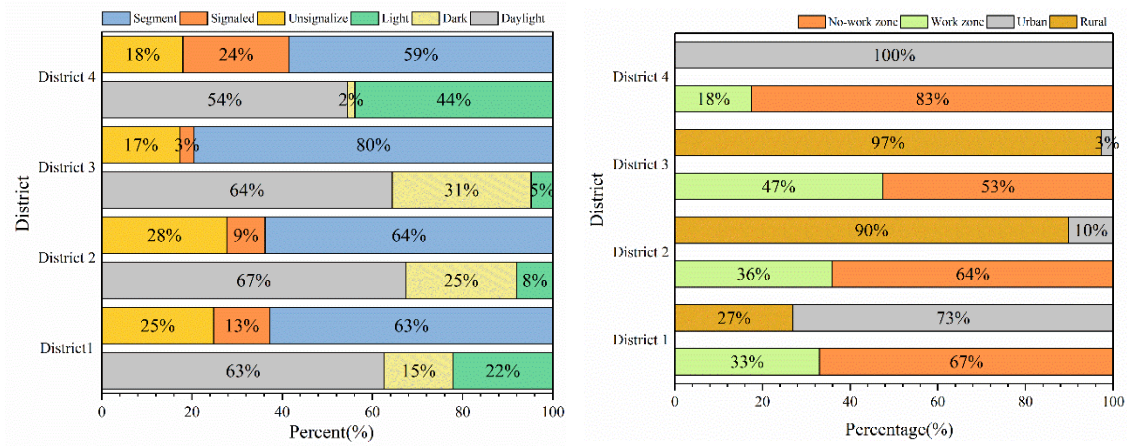

(a) Intersection and light condition  (b) Work zones and land use

**Fig.3** The spatial distribution characteristics of the crashes in four districts

Note: Segment represents crash occurred on the road segment; Signalized represents signalized intersections; Unsignalized represents Unsignalized intersections.

**Table 1 Description statistics**

| Variables | | District 1 | | District 2 | | District 3 | | District 4 | |
|---|---|---|---|---|---|---|---|---|---|
| | | Means | S.D. | Means | S.D. | Means | S.D. | Means | S.D. |
| No injury | Total number of the speeding-related crashes with no injury | 357 | | 796 | | 447 | | 1613 | |
| Minor injury | Total number of the speeding-related crashes with minor injury | 172 | | 367 | | 118 | | 1582 | |
| Serious injury | Total number of the speeding-related crashes with serious injury | 47 | | 103 | | 69 | | 238 | |
| **Crash Characteristics** | | | | | | | | | |
| Rear-end | 1 if the crash was recorded as a rear-end crash, 0 otherwise | .191 | .393 | .217 | .411 | .124 | .329 | .366 | .482 |
| Angle crash | 1 if the crash was recorded as a angle crash, 0 otherwise | .165 | .371 | .134 | .341 | .051 | .220 | .203 | .402 |
| Head-on | 1 if the crash was recorded as a head-on crash, 0 otherwise | .036 | .187 | .019 | .136 | .029 | .170 | .029 | .167 |
| Sideswipe | 1 if the crash was recorded as a sideswipe crash, 0 otherwise | .047 | .212 | .027 | .162 | .016 | .124 | .082 | .274 |
| Hit fixed object | 1 if the crash was recorded as a hit fixed object crash, 0 otherwise | .526 | .499 | .551 | .498 | .720 | .449 | .286 | .452 |
| Hit pedestrian | 1 if the crash was recorded as a hit pedestrian crash, 0 otherwise | .005 | .072 | .002 | .040 | .003 | .053 | .019 | .135 |
| Hit bicycle | 1 if the crash was recorded as a hit bicycle crash, 0 otherwise | 0 | 0 | .001 | .028 | .001 | .038 | .003 | .057 |
| **Environment and Roadway characteristics** | | | | | | | | | |
| State road | 1 if the crash took place on an state road, 0 otherwise | .233 | .423 | .188 | .390 | 0 | 0 | .465 | .499 |
| Local road | 1 if the crash took place on a local road, 0 otherwise | .356 | .479 | .429 | .495 | .311 | .463 | .390 | .488 |
| Curve road | 1 if the crash occurred on a curve road segment, 0 otherwise | .330 | .471 | .359 | .480 | .474 | .499 | .175 | .380 |
| Speed limit | 1 if speed limit ≥50mile/h , 0 otherwise | .302 | .459 | .322 | .468 | .308 | .462 | .447 | .497 |
| Unsignalized intersection | 1 if the crash took place at an unsignalized intersection, 0 otherwise | .248 | .432 | .277 | .448 | .173 | .379 | .180 | .385 |
| Signalized intersection | 1 if the crash took place at a signalized intersection, 0 otherwise | .125 | .327 | .085 | .278 | .031 | .174 | .235 | .424 |
| Snow | 1 if the crash involved a snow or slush covered road, 0 otherwise. | .217 | .412 | .160 | .367 | .183 | .387 | .020 | .140 |
| Work | 1 if the crash occurred at work zone, 0 otherwise | .023 | .149 | .003 | .056 | .006 | .075 | .033 | .178 |
| Dark | 1 if the crashes occurred at night without lights, 0 otherwise. | .152 | .360 | .246 | .431 | .308 | .462 | .017 | .131 |

| | | | | | | | | |
|---|---|---|---|---|---|---|---|---|
| Light | 1 if the crashes occurred at night with lights, 0 otherwise. | .222 | .416 | .08 | .272 | .048 | .211 | .439 | .496 |
| Rural | 1 if crash occurs at rural area, 0 otherwise. | .270 | .444 | .898 | .303 | .973 | .162 | 0 | 0 |
| **Driver and vehicle characteristics** | | | | | | | | | |
| Older driver | 1 if the crash involved at least 1 driver age ≥60, 0 otherwise | .099 | .299 | .099 | .298 | .095 | .294 | .064 | .244 |
| Young driver | 1 if the crash involved at least 1 driver age ≤20, 0 otherwise | .179 | .384 | .238 | .426 | .288 | .453 | .141 | .348 |
| Drunk driving | 1 if the crash involved at least 1 drunk driver, 0 otherwise | .1233 | .3293 | .055 | .227 | .079 | .270 | .050 | .217 |
| Exceeding the speed limit | 1 if the crash involved at least 1 driver exceeded speed limit, 0 otherwise | .140 | .348 | .132 | .339 | .165 | .371 | .204 | .403 |
| Fatigued driving | 1 if the crash involved at least 1 driver with fatigued driving, 0 otherwise | .003 | .059 | .010 | .100 | .011 | .106 | .009 | .096 |
| Drug related | 1 if the crash involved at least 1 driver with drugs reported, 0 otherwise | .057 | .232 | .025 | .157 | .041 | .199 | .027 | .163 |
| Running stop sign | 1if the crash involved at least a driver ran a stop sign, 0 otherwise | .028 | .164 | .022 | .147 | .027 | .162 | .013 | .114 |
| Running red light | 1 if the crash involved at least 1 driver ran a red light, 0 otherwise | .014 | .117 | .012 | .112 | .006 | .075 | .020 | .140 |
| Unbelted | 1 if the crash involved at least 1 person unbelted, 0 otherwise | .135 | .342 | .103 | .305 | .136 | .343 | .148 | .355 |
| Large Truck | 1 if the crash involved at least 1 truck, 0 otherwise | .073 | .260 | .137 | .344 | .061 | .239 | .053 | .225 |
| Overturn | 1if the crash involved at least one overturned vehicle, 0 otherwise | .069 | .254 | .116 | .320 | .168 | .374 | .038 | .191 |

## 5. Spatial Stability Test

Likelihood ratio (LR) test is conducted to determine whether speeding-related crashes classified by severe crash cluster districts should be modeled separately. Firstly, the spatial stability of the parameter estimates between the two districts is tested by Eq. (14)

$$\chi^2 = -2\left[\text{LL}(\beta_{m_1,m_2}) - \text{LL}(\beta_{m_1})\right] \tag{14}$$

Where $\text{LL}(\beta_{m_1})$ represents the log-likelihood of a converged model using the data collected from district $m_1$, $\text{LL}(\beta_{m_1,m_2})$ denotes the log-likelihood values of a converged model using both parameters from $m_2$ and the data from district $m_1$, and $\chi^2$ is a test statistic (chi-square distributed) with degrees of freedom equal to the number of estimated parameters in $\beta_{m_1,m_2}$(Behnood and Mannering 2015). As shown in Tables 2, 9 of all 12 tests produce a confidence level more than 90% (except for $m_2/m_1$ for District 1/ District 2, District 1/District 3, and District 2/District 3 which produce less than 90% confidence level; however, the reversed tests of these models produce 90% confidence level), indicating that the null hypothesis is rejected at a high confidence level(Yan et al. 2021a). Therefore, it can be concluded that the parameters are spatial instability between the two districts.

**Table 2 Likelihood ratio test results between different districts**

| m1 | m2 | | | |
|---|---|---|---|---|
| | District 1 | District 2 | District 3 | District 4 |
| District 1 | -- | 17.40(18) [50.4%] | 14.88(15) [53.9%] | 38.62(17) [99.7%] |
| District 2 | 43.92(16) [>99.9%] | -- | 13.04(15) [40.0%] | 47.21(18) [>99.9%] |
| District 3 | 24.46(16) [92.1%] | 46.34(18) [90.99%] | -- | 52.64 (18) [>99.9%] |
| District 4 | 306.38(16) [>99.9%] | 183.72(18) [>99.9%] | 106.7(15) [>99.9%] | -- |

The second likelihood ratio (LR) test is employed between the full model and the four sub-models to further investigate whether speeding-related crashes should be modelled separately at four districts, as shown in Eq. (15).

$$\chi^2_{full} = -2\left[\text{LL}(\beta_{full}) - \sum_{\alpha=1}^{4}\text{LL}(\beta_{D\alpha})\right] \tag{15}$$

Where $\chi^2_{full}$ represents the log-likelihood of a converged model using the speeding-related crash data from four districts, $\text{LL}(\beta_{D\alpha})$ represents the log-likelihood of a converged model using the crash data from district α. $\chi^2_{full}$ follows a chi-square distribution with degrees of freedom equal to the summation of parameters found to be statistically significant in each district minus the number of parameters found to be statistically significant using the data from four districts at the same time. The second LR test result is 234.2 with 25 degrees of freedom suggesting that the null hypothesis can be rejected. Therefore, it can be concluded that speeding related crashes should be modelled separately for four severe crash cluster districts.

## 6 Results

As described in Section 4, the four data subsets are first extracted from hotspot districts. The CRPM-Probit model is used to capture unobserved heterogeneity. All random parameters are assumed to follow a normal distribution, and 1000 Halton draws are employed. Only variables with at least 90% confidence in statistical significance are retained. The primary aim of this research paper was to establish four discrete CRPM-Probit models, with the purpose of investigating the disparate factors that influence the severity of speeding-related crashes across four distinct districts. The analysis, however, has revealed that no statistically significant factors could be discerned to exert an influence on the mean of random parameters in the models pertaining to Districts 1 and 3. Hence, the models in these two districts have devolved into CRP-Probit models. In order to ascertain the validity of the model, five distinct models, namely Probit, RP-Probit, RPM-Probit, CRP-Probit, and CRPM-Probit models, have been chosen for comparative analysis. Two indicators, McFadden Pseudo $R^2$(Pseudo $R^2$) and Akaike information criterion (AIC), are employed to evaluate the overall goodness-of-fit of the estimated models (Fanyu et al,2021). The model comparison results are shown in Table 3.

Table 3 Goodness of fit measures

| Model | District 1 | | District 2 | | District 3 | | District 4 | |
|---|---|---|---|---|---|---|---|---|
| | Pseudo $R^2$ | AIC | Pseudo $R^2$ | AIC | Pseudo $R^2$ | AIC | Pseudo $R^2$ | AIC |
| Probit | .108 | 922.3 | .083 | 2015.1 | .118 | 1114.2 | .095 | 5609.8 |
| RP-Probit | .110 | 907.8 | .086 | 2014.6 | .121 | 1109.3 | .096 | 5600.1 |
| RPM-Probit | -- | -- | .086 | 2015.8 | -- | -- | .096 | 5598.3 |
| CRP-Probit | .121 | 904.5 | .086 | 2016.5 | .122 | 1103.8 | .096 | 5598.7 |
| CRPM-Probit | -- | -- | .088 | 2014.3 | -- | -- | .096 | 5597.9 |

In District 1 and 3, the CRP-Probit model had the highest Pseudo $R^2$ and the lowest AIC values. The result provides empirical evidence to support the superiority of the CRP-Probit models in terms of performance when compared to both the RP-Probit and Probit models. In District 2 and District 4, the CRPM-Probit model exhibited the lowest AIC values of 2014.3 and 5597.9, respectively. This observation implied that the CRPM-Probit model could have induced less information loss by incorporating the possible correlations between the random parameters. Therefore, the CRPM-Probit and CRP-Probit models and were the most effective in analyzing the severity of speeding-related crashes in these districts. The parameters' estimate, Akaike information criterion (AIC), t-statistics, goodness-of-fit statistics, likelihood at convergence, and marginal effect are presented in Table 4.

**6.1 Crash Characteristics**

Regarding crash characteristics, differences among the significant variables associated with crash severity across the four districts can be observed. Rear-end crashes are significantly positively associated with increased serious injury (by 0.9%) in District 2 but negatively correlated (by 3.4%) in District 3. Sideswipe crashes result in a lower likelihood of serious injury in District 4 (by 1.2%) and present a 1.7% increase in District 2. The results further illustrate that the effect of collision characteristics on crash severity is spatial instability. Rear-end crashes, angle crashes, head-on crashes, and sideswipe crashes are found to increase crash severity in District 2, indicating that more traffic management measures should be implemented in this district that aim to mitigate multi-vehicle (at least two vehicles involved) crash severity. Hit pedestrian crashes are significantly positively associated with increased serious injury (23.7%), but only in District 4. Therefore, improving pedestrian safety is a priority in District 4. In Districts 1 and 3, crash characteristics are not significantly related to increase crash severity, and more attention should be given to other factors.

**6.2 Roadway and environment characteristics**

In terms of the roadway characteristics, a speed limit above 50 mi/h in District 4 is associated with a lower probability of serious injury (reduction by 2.3%). The results indicate speeding vehicles on low speed limit road segments in District 4 should receive more attention than those with speed limits above 50mi/h. Speeding-related crashes at unsignalized intersection increased the probability of serious injury crashes by 0.14% in District 1, but heterogeneity was captured in District 2. Signalized intersections are significantly positively associated with increased crash severity in District 4, but negatively associated in Districts 2 and 3. These results suggest that the relationship between crash severity and intersections varies by district. The work zone variable is an influential factor in District 2 and increases the risk of serious injury by 27.6%.

Regarding the environment characteristics, light conditions have been found to play an essential role as a key risk factor influencing speeding-related crash injury severity in Districts 1 and 3. The dark indicator is associated with more severe injuries in District 1. Speeding-related crashes occurring at the night with streetlamps tend to decrease the likelihood of serious injury in Districts 1 and 3. The rural area indictor is significant in District 1, which increases the likelihood of serious injury by 0.1%.

**6.3 Driver and vehicle characteristics**

Regarding driver characteristics, drivers younger than 20 years of age are associated with serious injuries in District 1 but have random effects in Districts 3 and 4. In District 2, older drivers increase the likelihood of serious- and minor-injury crashes by 0.8% and 15.7%. This result indicates that the age-related indicators vary by district.

The model results also provided evidence of priority districts for traffic enforcement. Driver status during the crash, including drunk driving, speeding (exceeding the speed limit), and not wearing a seatbelt, are significantly positively associated with increased crash severities in the four districts. Fatigued driving increases the likelihood of serious injury crashes by 5.7% in District 2. Drug-related crashes are found to be significant in Districts 3 and 4. The behavior of running stop signs highly increased crash severity in District 2. Running red lights increased the possibility of serious injury by 11.6% in District 1. The model results did not uncover vehicle type differences in crash injury severity. However, overturning increased crash severity in Districts 2, 3, and 4.

**6.4 Heterogeneity**

Among the various crash, roadway, environment, driver, and vehicle factors examined, a total of five variables shows significant standard deviations, indicating the presence of heterogeneity. For instance, in District 1, a snow or slush road is statistically significant with a mean of -.808 and a standard deviation of .993. The results indicate that a snow- or slush-covered road can reduce the likelihood of serious injury in 65.54% of observations while increasing injury severity in 34.46% of observations. This finding is understandable, as many observed and unobserved factors, such as the road surface type (fresh snow, compacted snow, and slush), snowfall, and traffic volume, have a complex effect on the coefficient of friction(Yasmin and Eluru 2013). The distribution effect of the random parameters across observations in the four districts is presented in Table 3.

Heterogeneity in the means of random parameters is calculated to further investigate the impacts of the random coefficients. The model results show that intercept is positively associated with unsignalized intersections in District 2, indicating that speeding-related crashes at unsignalized intersections are more dangerous than in other road segments. A curve road increases the mean of speeding driving (exceeding the speed limit) in District 4. This suggests that driving above the speed limit on a curved road increases crash severity.

In addition, correlations between random parameters are also observed in the four districts. Intercepts are positively correlated with the random parameter of "*Snow*" (.9611) in District 1, indicating that a snow- or slush-covered covered road is likely to cause severe crash outcomes. In District 2, a negative correlation (-.6127) between intercept and light indicates that speeding-related crashes occurring at night with streetlamps reduce the likelihood of serious injury. Signaled intersection and exceeding the speed limit are found to be negatively correlated with young drivers in District 3 and 4 respectively.

**Table 4 Results of estimated parameters and marginal effects**

| Variables | District 1 | | | | | District 2 | | | | | District 3 | | | | | District 4 | | | | |
|---|---|---|---|---|---|---|---|---|---|---|---|---|---|---|---|---|---|---|---|---|
| | Cor. | t-stat | Margnal effect | | | Cor. | t-stat | Margnal effect | | | Cor. | t-stat | Margnal effect | | | Cor. | t-stat | Margnal effect | | |
| | | | S | M | N | | | S | M | N | | | S | M | N | | | S | M | N |
| Constant | -.893 | -6.01 | | | | -1.058 | -12.32 | | | | -.131 | -1.07 | | | | .041 | .78 | | | |
| S.D | 1.730 | 13.57 | | | | 1.350 | 18.79 | | | | | | | | | | | | | |
| **Crash Characteristics** | | | | | | | | | | | | | | | | | | | | |
| Rear-end | | | | | | .553 | 4.78 | .009 | .187 | -.196 | -.390 | -2.20 | -.034 | -.099 | .134 | | | | | |
| Angle crash | | | | | | .563 | 4.06 | .010 | .193 | -.204 | | | | | | .192 | 3.10 | .017 | .057 | -.075 |
| Head-on | | | | | | 1.288 | 4.38 | .071 | .410 | -.481 | | | | | | | | | | |
| Sideswipe | | | | | | .670 | 2.73 | .017 | .235 | -.251 | | | | | | -.163 | -2.04 | -.012 | -.053 | .065 |
| Hit fixed object | -1.003 | -6.59 | -.002 | -.308 | .311 | | | | | | -.643 | -5.12 | -.091 | -.153 | .244 | -.339 | -6.60 | -.025 | -.109 | .134 |
| Hit pedestrian | | | | | | | | | | | | | | | | 1.191 | 5.67 | .237 | .127 | -.365 |
| **Roadway and environment characteristics** | | | | | | | | | | | | | | | | | | | | |
| Speed limit | | | | | | | | | | | | | | | | -.275 | -5.28 | -.023 | -.086 | .108 |
| Unsignalized intersection | .541 | 3.36 | .001 | .181 | -.182 | -.340 | -2.80 | -.003 | -.103 | .106 | | | | | | | | | | |
| S.D | | | | | | 1.530 | 12.94 | | | | | | | | | | | | | |
| Signalized intersection | | | | | | -.782 | -4.10 | -.004 | -.194 | .198 | -.459 | -1.30 | -.036 | -.115 | .151 | .297 | 5.14 | .029 | .087 | -.116 |
| S.D | | | | | | | | | | | 1.045 | 2.46 | | | | | | | | |
| Snow | -.808 | -3.04 | -.001 | -.207 | .208 | -.523 | -3.91 | -.003 | -.147 | .150 | | | | | | | | | | |
| S.D | .993 | 4.94 | | | | | | | | | | | | | | | | | | |
| Work zone | | | | | | 2.133 | 2.75 | .276 | .393 | -.669 | | | | | | | | | | |
| Dark | .353 | 1.70 | .001 | .117 | -.118 | | | | | | | | | | | | | | | |
| Light | -.352 | -1.81 | -.001 | -.101 | .102 | | | | | | -.409 | -1.68 | -.034 | -.103 | .137 | | | | | |
| Rural | .437 | 2.56 | .002 | .143 | -.145 | | | | | | | | | | | | | | | |

| | District 1 | | | | | District 2 | | | | | District 3 | | | | | District 4 | | | | |
|---|---|---|---|---|---|---|---|---|---|---|---|---|---|---|---|---|---|---|---|---|
| **Driver and vehicle characteristics** | | | | | | | | | | | | | | | | | | | | |
| Older driver | | | | | | .458 | 3.25 | .008 | .157 | -.165 | | | | | | | | | | |
| Young driver | .938 | 5.19 | .005 | .330 | -.335 | | | | | | -.249 | -2.18 | -.025 | -.064 | .090 | -.085 | -1.41 | -.006 | -.027 | .034 |
| S.D | | | | | | | | | | | .270 | 2.82 | | | | .468 | 7.99 | | | |
| Drunk driving | .942 | 4.34 | .0058 | .337 | -.342 | 1.619 | 9.60 | .116 | .463 | -.579 | .794 | 4.24 | .145 | .162 | -.308 | .386 | 4.45 | .044 | .104 | -.147 |
| Exceeding the speed limit | 1.047 | 5.37 | .0072 | .373 | -.381 | .718 | 6.02 | .016 | .247 | -.263 | .362 | 2.89 | .049 | .089 | -.138 | .282 | 4.83 | .027 | .082 | -.110 |
| S.D | | | | | | | | | | | | | | | | .606 | 12.53 | | | |
| Asleep or Fatigued | | | | | | 1.170 | 2.95 | .057 | .384 | -.441 | | | | | | | | | | |
| Drug related | | | | | | | | | | | .809 | 3.77 | .155 | .159 | -.314 | .236 | 2.10 | .024 | .067 | -.091 |
| Running stop sign | | | | | | .538 | 1.96 | .011 | .187 | -.198 | | | | | | | | | | |
| Running red light | 2.173 | 3.64 | .116 | .577 | -.693 | | | | | | | | | | | | | | | |
| Unbelted | 1.508 | 7.83 | .021 | .522 | -.543 | .896 | 6.96 | .026 | .309 | -.334 | 1.090 | 8.06 | .216 | .197 | -.414 | .511 | 8.88 | .058 | .135 | -.193 |
| Overturn | | | | | | .780 | 6.05 | .019 | .269 | -.289 | .325 | 2.63 | .043 | .081 | -.124 | .569 | 5.84 | .074 | .135 | -.209 |
| **Heterogeneity in means** | | | | | | | | | | | | | | | | | | | | |
| Constant: Un-intersection | | | | | | .947 | 3.46 | | | | | | | | | | | | | |
| Exceeding-limit: Curve road | | | | | | | | | | | | | | | | .337 | 2.36 | | | |
| **Threshold parameters** | | | | | | | | | | | | | | | | | | | | |
| $u_1$ | 2.631 | 13.33 | | | | | | | | | 1.20 | 14.54 | | | | 1.873 | 43.23 | | | |
| **Distribution of random parameters** | | | | | | | | | | | | | | | | | | | | |

| | District 1 | | District 2 | | District 3 | | District 4 | |
|---|---|---|---|---|---|---|---|---|
| | Above zero | Below zero | Above zero | Below zero | Above zero | Below zero | Above zero | Below zero |
| Snow | 34.46% | 65.54% | | | | | | |
| Unsignaled intersection | | | 41.29% | 58.71% | | | | |
| Signaled intersection | | | | | 33% | 67% | | |
| Young driver | | | | | 17.88% | 82..12% | 42.86% | 57.14% |

| | | | | | | | |
|---|---|---|---|---|---|---|---|
| Exceeding the speed limit | | | | | | 68.08% | 31.92% |

**Diagonal and off-diagonal matrix[t-stats], and correlation coefficients (in parenthesis) of random parameters**

| | **District 1** | | **District 2** | | **District 3** | | **District 4** | |
|---|---|---|---|---|---|---|---|---|
| Random parameter | Snow | | Light | | Signaled intersection | | Young driver | |
| Constant | 3.456[11.42]( .9611) | Constant | -1.186 [-11.03]( -.6127) | Young driver | -.541[-5.26]( -.565) | Exceeding the speed limit | -.110[-1.91]( -.229) | |

**Model statistics**

| | | District 1 | District 2 | District 3 | District 4 |
|---|---|---|---|---|---|
| Number of observations | | | 1266 | | 3433 |
| Number of estimated parameters | | 16 | 18 | 15 | 18 |
| LL (0) | Log-Likelihood (only constant) | -496.44 | -1082.21 | -611.52 | -3079.21 |
| LL (β) | Likelihood at convergence | -436.25 | -986.15 | -536.89 | -2780.96 |
| ρ²=1 - LL(β)/LL(0) | | .121 | .088 | .122 | .096 |
| AIC | | 904.5 | 2014.3 | 1103.8 | 5597.9 |

**Note :** Variables defined for N = no injury, M= minor injury, N= no injury, Cor.= coefficient, S.D= Standard deviation.

# 7. Discussion

From the crash perspective, the model findings reveal that vehicle-vehicle crashes increase crash severity in District 2, and the head-on collision type tends to be the most dangerous. This finding is logical and consistent with expectations(Hosseinpour et al. 2014; Lee and Li 2014; Liu and Fan 2020). Pedestrians are considered the most vulnerable traffic participants (Dai 2012). Hit pedestrian crash is found to be the most dangerous collision type in District 4. A possible explanation is that pedestrians are directly exposed to crashes(Huang et al. 2008; Li and Fan 2018), and speeding vehicles significantly increase pedestrian injury severity. These results indicate that management measures should vary by district. Multi-vehicle crashes (at least two vehicles involved) should be given priority attention in District 2. In contrast, improving pedestrian safety is more important in District 4.

From the roadway perspective, a higher speed limit (≥50mi/h) reduces speeding-related crash severity in District 4. This finding contradicts previous studies, which posit that higher speed limits could adversely affect traffic safety (Uddin and Huynh 2020; Yasmin and Eluru 2013). As mention in Fig.3, district 4 is urban areas. According to (Demiroz et al. 2015)Demiroz et al. (2015), pedestrians perceive that crossing roads with higher speed limits is riskier than those with lower speed limits, which might result in a decrease in the frequency of illegal road crossing behavior in areas with higher speed limits in District 4. Speeding at unsignalized and signaled intersections increased crash severity in Districts 1 and 4, respectively, but heterogeneity was captured in Districts 2 and 3. This result is complex and may be caused by the unobserved factors in crash data, such as intersection geometric features, topographical features, and traffic flow characteristics. This finding suggests that the traffic management department should focus on improving the safety of unsignalized intersections in District 1, and on signaled intersections in District 4.

From the environmental perspective, the work zone variable is influential in District 2 but insignificant in other districts. This may be due to the differences in safety regulations at the work zone, as well as the size of the work zone, and the duration of construction(Lym and Chen 2021). Regarding the effect of lighting conditions, speeding–related crashes occurring at night without streetlamps increase the probabilities of serious injury, while the severity of crashes is reduced in lighted conditions. Poor driver visibility in dark conditions is a major cause of serious crashes(Anarkooli et al. 2017; Jalayer et al. 2018). Drivers are more likely to be faced with unexpected events under dark conditions, and it is difficult for them to identify the risk of a crash quickly. This finding suggests that it is necessary to improve road lighting conditions in District 1. The rural area indicator significantly increases crash severity compared to those that occurred in urban areas in District 1. The higher frequency of vehicle speeding can explain this result due to the lower traffic volume in rural areas.

From the driver and vehicle characteristics perspectives, driver age is a significant factor affecting crash severity. Older drivers are associated with more severe crash outcomes in District 2. Cognitive limitations and physical abilities make older drivers more likely to be involved in serious crashes, especially when faced with speeding-related crashes(Hanrahan et al. 2009; Islam and Burton 2019). Young drivers are inexperienced but physically strong, leading to the correlation between young drivers and crash severity (especially those related to speeding) varying geographically (Chen et al. 2016a; Lee and Li 2014). The model results confirmed previous findings that the probability of more severe consequences increases with drunk driving(Adanu et al. 2021; Shyhalla 2014; Zhang and Hassan 2019a), driving exceeding the speed limit(Se et al. 2021a; Se et al. 2021b), not wearing a seatbelt(Azimi et al. 2020; Yasmin and Eluru 2013), and overturning(Peng et al. 2018). At the same time, some driving behaviors are spatially unstable, such as being asleep or fatigued driving, drug-related driving, running stop signs and running red lights. These findings suggest priority areas for traffic enforcement.

**8. Conclusion and recommendations**

The present study focuses on investigating and analyzing variations in the effect of contributing factors on speeding-related crash severity across different severe crash cluster districts. The Global Moran's I coefficient shows a spatial clustering pattern within the state of Pennsylvania. The local Getis–Ord G* index is then used to identify severe crash cluster districts, and four crash datasets are extracted from four hotspot districts. Two LR tests are conducted to validate whether speeding-related crashes classified by crash cluster districts should be modeled separately. The results suggest that separate modeling is necessary. Four CRPO-Probit models with heterogeneity in means are applied to explore the heterogeneity in crash injury severity outcomes. According to the estimation results, the study provides recommendations to mitigate speeding-related crash severity from the engineering, technology, education, and enforcement perspectives.

From the engineering and technology perspective, improving traffic safety at unsignalized intersections and in dark conditions (without streetlamps) is a priority for District 1. Some traffic management strategies can be considered for implementation, such as improving the visibility of intersections by providing enhanced signing (providing lighting), closing or relocating "High-Risk" intersections, increasing the frequency of road lighting maintenance, and promoting internally illuminated LED traffic sign lights (Shauna Hallmark, 2020). Vehicle-vehicle crashes, work zone, and older drivers significantly contribute to the increased crash severity in District 2. Our study suggests that a unique design consideration is needed for undivided roadways to prevent head-on crashes, where reduced intersection skew angles can be used to prevent angle crashes in District 2. Work zones and special signings, such as those signs used at special events to direct traffic, must be maintained and used correctly. Improving the roadway and driving environment to better accommodate older drivers' unique needs is necessary for District 2. Some measures for improvement include increasing the size and letter height of roadway signs, replacing painted channelization with raised channelization, and implementing wider edge lines(Park et al. 2020). Traffic management in District 4 should give priority to vehicle-pedestrian crashes. Advanced traffic management technology can alleviate this problem by implementing alert system for drivers based on traffic signs, lights, and pedestrian detection.

From the enforcement perspective, this study suggests district priorities for traffic enforcement. For instance, speed enforcement should be prioritized at unsignalized intersections in District 1, while work zones and signalized intersections are paramount in District 2. The traffic management department should focus on the governance of running a red light in District 1. However, prioritizing reducing the running of stop signs and asleep or fatigued driving behaviors would be beneficial for District 2. Our results indicate that drug-related driving is a significant factor in Districts 3 and 4. Therefore, it is necessary to increase the frequency of random law enforcement against drug-related drivers in Districts 3 and 4. Drunk driving and being unbelted are associated with a higher crash severity in all four districts. Stricter penalties for drunk driving, such as driver's license suspension, increased jail time, and installing an ignition interlock device, could effectively reduce the frequency of drunk-driving. From the education perspective, the traffic management department in District 4 should improve pedestrian safety awareness via more effective ways, such as through awareness campaigns, media. There are also some limitations are to be acknowledged in this paper. Firstly, the temporal instability is ignored. Future research may consider spatiotemporal stability in the analysis of speeding-related crash injury severity. Second, CRPO-Probit models ignore the heterogeneity in threshold and variance heterogeneity. Future research may consider relax these restrictions to achieve accurate results. Third, the extraction of current traffic accident samples can be influenced to some extent by the personnel involved in the process. ChatGPT has started to generate utility in the fields of medicine and transportation (Huang et al.,2023; Zheng et al.,2023). In the future, it can be considered to utilize ChatGPT for conducting relevant research in this area.

**Acknowledgments**

The authors declare that there is no conflict of interest about this manuscript. This research is funded by the National Natural Science Foundation of China (No. 71871059), Postgraduate Research & Practice innovation Program of Jiangsu Province (No.KYCX22_0270), and China Scholarship Council (CSC)

**Reference:**

Aarts, L., & van Schagen, I. (2006). Driving speed and the risk of road crashes: a review. *Accid Anal Prev, 38*(2), 215-224.

Adanu, E. K., Agyemang, W., Islam, R., & Jones, S. (2021). A comprehensive analysis of factors that influence interstate highway crash severity in Alabama. *Journal of Transportation Safety & Security*, 1-25.

Afghari, A.P., Haque, M.M. and Washington, S., 2020. Applying a joint model of crash count and crash severity to identify road segments with high risk of fatal and serious injury crashes. Accident Analysis & Prevention, 144, p.105615.Alarifi, S.A., Abdel-Aty, M., Lee, J., 2018. A Bayesian multivariate hierarchical spatial joint model for predicting crash counts by crash type at intersections and segments along corridors. Accident Analysis & Prevention, 119, pp. 263-273.

Alrejjal, A., Moomen, M., & Ksaibati, K. (2022). Evaluating the effectiveness of law enforcement in reducing truck crashes for a rural mountainous freeway in Wyoming. Transportation letters, 14(8), 807-817.

Ahie, L. M., Charlton, S. G., & Starkey, N. J. (2015). The role of preference in speed choice. *Transportation Research Part F: Traffic Psychology and Behaviour, 30*, 66-73.

Alrejjal, A., Moomen, M., & Ksaibati, K. (2022). Evaluating the impact of traffic violations on crash injury severity on Wyoming interstates: An investigation with a random parameters model with heterogeneity in means approach. *Journal of Traffic and Transportation Engineering (English Edition)*.

Anarkooli, A. J., Hosseinpour, M., & Kardar, A. (2017). Investigation of factors affecting the injury severity of single-vehicle rollover crashes: A random-effects generalized ordered probit model. *Accid Anal Prev, 106*, 399-410.

Azimi, G., Rahimi, A., Asgari, H., & Jin, X. (2020). Severity analysis for large truck rollover crashes using a random parameter ordered logit model. *Accident Analysis & Prevention, 135*.

Behnood, A., & Mannering, F. (2019). Time-of-day variations and temporal instability of factors affecting injury severities in large-truck crashes. *Analytic Methods in Accident Research, 23*.

Behnood, A., & Mannering, F. L. (2015). The temporal stability of factors affecting driver-injury severities in single-vehicle crashes: Some empirical evidence. *Analytic Methods in Accident Research, 8*, 7-32.

Bhowmik T., M. Rahman., S. Yasmin and N. Eluru (2021). "Exploring Analytical, Simulation-Based, And Hybrid Model Structures For Multivariate Crash Frequency Modeling", Analytic Methods in Accident Research Volume 31, September 2021, 100167

Bhowmik T., S. Yasmin and N. Eluru (2018), "A Joint Econometric Approach for Modeling Crash Counts by Collision Type", Analytic Methods in Accident Research Volume 19, September 2018, Pages 16-32

Chen, C., Zhang, G., Huang, H., Wang, J., & Tarefder, R. A. (2016). Examining driver injury severity outcomes in rural non-interstate roadway crashes using a hierarchical ordered logit model. *Accid Anal Prev, 96*, 79-87.

Chen, C., Zhang, G., Liu, X. C., Ci, Y., Huang, H., Ma, J., . . . Guan, H. (2016). Driver injury severity outcome analysis in rural interstate highway crashes: a two-level Bayesian logistic regression interpretation. *Accid Anal Prev, 97*, 69-78.

Dai, D. (2012). Identifying clusters and risk factors of injuries in pedestrian–vehicle crashes in a GIS environment. *Journal of Transport Geography, 24*, 206-214.

Demiroz, Y. I., Onelcin, P., & Alver, Y. (2015). Illegal road crossing behavior of pedestrians at overpass locations: Factors affecting gap acceptance, crossing times and overpass use. *Accid Anal Prev, 80*, 220-228.

Eluru, Naveen, Chandra R. Bhat, and David A. Hensher. "A mixed generalized ordered response model for examining pedestrian and bicyclist injury severity level in traffic crashes." Accident Analysis & Prevention 40.3 (2008): 1033-1054.

Mannering, F.L., Shankar, V. and Bhat, C.R., 2016. Unobserved heterogeneity and the statistical analysis of highway accident data. Analytic methods in accident research, 11, pp.1-16.

Fountas, Grigorios, et al. "Addressing unobserved heterogeneity in the analysis of bicycle crash injuries in Scotland: A correlated random parameters ordered probit approach with heterogeneity in means." Analytic methods in accident research 32 (2021): 100181.

Feng, S., Li, Z., Ci, Y., & Zhang, G. (2016). Risk factors affecting fatal bus accident severity: Their impact on different types of bus drivers. *Accident Analysis & Prevention, 86*, 29-39.

Fanyu, M., Sze, N. N., Cancan, S., Tiantian, C., & Yiping, Z. (2021). Temporal instability of truck volume composition on non-truck-involved crash severity using uncorrelated and correlated grouped random parameters binary logit models with space-time variations. Analytic Methods in Accident Research, 31, 100168.

Getis, A., & Ord, J. K. (2010). The analysis of spatial association by use of distance statistics. In Perspectives on spatial data analysis (pp. 127-145). Springer, Berlin, Heidelberg.

Guo, Y., Li, Z., Liu, P. and Wu, Y., 2019. Modeling correlation and heterogeneity in crash rates by collision types using full Bayesian random parameters multivariate Tobit model. Accident Analysis & Prevention, 128, pp.164-174.Hanrahan, R. B., Layde, P. M.,

Zhu, S., Guse, C. E., & Hargarten, S. W. (2009). The association of driver age with traffic injury severity in Wisconsin. *Traffic Inj Prev, 10*(4), 361-367.

Hosseinpour, M., Yahaya, A. S., & Sadullah, A. F. (2014). Exploring the effects of roadway characteristics on the frequency and severity of head-on crashes: case studies from Malaysian federal roads. *Accid Anal Prev, 62*, 209-222.

Hoye, A. (2020). Speeding and impaired driving in fatal crashes-Results from in-depth investigations. *Traffic Inj Prev, 21*(7), 425-430.

Hou, Q., Huo, X., Leng, J., & Cheng, Y. (2019). Examination of driver injury severity in freeway single-vehicle crashes using a mixed logit model with heterogeneity-in-means. Physica A: Statistical Mechanics and its Applications, 531.

Huang, H., Abdel-Aty, M. A., & Darwiche, A. L. (2010). County-Level Crash Risk Analysis in Florida: Bayesian Spatial Modeling. *Transportation Research Record: Journal of the Transportation Research Board, 2148*(1), 27-37.

Huang, H., Chin, H. C., & Haque, M. M. (2008). Severity of driver injury and vehicle damage in traffic crashes at intersections: a Bayesian hierarchical analysis. *Accid Anal Prev, 40*(1), 45-54.

Huang, Y., Sun, D. J., & Tang, J. (2018). Taxi driver speeding: Who, when, where and how? A comparative study between Shanghai and New York City. *Traffic Inj Prev, 19*(3), 311-316.

Islam, M., & Mannering, F. (2021). The role of gender and temporal instability in driver-injury severities in crashes caused by speeds too fast for conditions. *Accid Anal Prev, 153*, 106039.

Islam, S., & Burton, B. (2019). A comparative injury severity analysis of rural intersection crashes under different lighting conditions in Alabama. *Journal of Transportation Safety & Security, 12*(9), 1106-1127.

Jalayer, M., Shabanpour, R., Pour-Rouholamin, M., Golshani, N., & Zhou, H. (2018). Wrong-way driving crashes: A random-parameters ordered probit analysis of injury severity. *Accid Anal Prev, 117*, 128-135.

Kabli, A., Bhowmik, T., and Eluru, N., 2020. A multivariate approach for modeling driver injury severity by body region. Analytic methods in accident research, 28, 100129

Lee, C., & Li, X. (2014). Analysis of injury severity of drivers involved in single- and two-vehicle crashes on highways in Ontario. *Accid Anal Prev, 71*, 286-295.

Lemp, Jason D., Kara M. Kockelman, and Avinash Unnikrishnan. "Analysis of large truck crash severity using heteroskedastic ordered probit models." Accident Analysis & Prevention 43.1 (2011): 370-380.

Li, Y., Abdel-Aty, M., Yuan, J., Cheng, Z., & Lu, J. (2020). Analyzing traffic violation behavior at urban intersections: A spatio-temporal kernel density estimation approach using automated enforcement system data. *Accid Anal Prev, 141*, 105509.

Li, Y., & Fan, W. (2018). Modelling the severity of pedestrian injury in pedestrian—vehicle crashes in North Carolina: A partial proportional odds logit model approach. *Journal of Transportation Safety & Security, 12*(3), 358-379.

Li, Z., Wang, W., Liu, P., Bigham, J. M., & Ragland, D. R. (2013). Using Geographically Weighted Poisson Regression for county-level crash modeling in California. *Safety Science, 58*, 89-97.

Liu, P., & Fan, W. (2020). Analyzing injury severity of rear-end crashes involving large trucks using a mixed logit model: A case study in North Carolina. *Journal of Transportation Safety & Security, 14*(5), 723-736.

Lym, Y., & Chen, Z. (2021). Influence of built environment on the severity of vehicle crashes caused by distracted driving: A multi-state comparison. *Accid Anal Prev, 150*, 105920.

Mannering, F. L., Shankar, V., & Bhat, C. R. (2016). Unobserved heterogeneity and the statistical analysis of highway accident data. *Analytic Methods in Accident Research, 11*, 1-16.

Mohamad, F. F., Abdullah, A. S., & Mohamad, J. (2019). Are sociodemographic characteristics and attitude good predictors of speeding behavior among drivers on Malaysia federal roads? *Traffic Inj Prev, 20*(5), 478-483.

NHTSA. United States Department of Transportation. Retrieved from https://www.nhtsa.gov/risky-driving/speeding

Osman, M., Mishra, S. and Paleti, R., 2018. Injury severity analysis of commercially licensed drivers in single-vehicle crashes: Accounting for unobserved heterogeneity and age group differences. Accident Analysis & Prevention, 118, pp.289-300.

Park, H. C., Yang, S., Park, P. Y., & Kim, D. K. (2020). Multiple membership multilevel model to estimate intersection crashes. *Accid Anal Prev, 144*, 105589.

Peng, Y., Wang, X., Peng, S., Huang, H., Tian, G., & Jia, H. (2018). Investigation on the injuries of drivers and copilots in rear-end crashes between trucks based on real world accident data in China. *Future Generation Computer Systems, 86*, 1251-1258.

Plug, C., Xia, J. C., & Caulfield, C. (2011). Spatial and temporal visualisation techniques for crash analysis. *Accid Anal Prev, 43*(6), 1937-1946.

Pulugurtha, S. S., Krishnakumar, V. K., & Nambisan, S. S. (2007). New methods to identify and rank high pedestrian crash zones: an illustration. *Accid Anal Prev, 39*(4), 800-811.

Rahman, M. H., Zafri, N. M., Akter, T., & Pervaz, S. (2021). Identification of factors influencing severity of motorcycle crashes in Dhaka, Bangladesh using binary logistic regression model. *Int J Inj Contr Saf Promot, 28*(2), 141-152.

Rezapour, M., & Ksaibati, K. (2022). Contributory factors to the severity of single-vehicle rollover crashes on a mountainous area, generalized additive model. *Int J Inj Contr Saf Promot*, 1-8.

Rezapour, M., Moomen, M., & Ksaibati, K. (2019). Ordered logistic models of influencing factors on crash injury severity of single and multiple-vehicle downgrade crashes: A case study in Wyoming. *J Safety Res, 68*, 107-118.

Rezapour, Mahdi, et al. "Application of Bayesian ordinal logistic model for identification of factors to traffic barrier crashes: Considering roadway classification." Transportation letters 13.4 (2021): 308-314.

Sadri, Arif Mohaimin, Satish V. Ukkusuri, and Pamela Murray-Tuite. "A random parameter ordered probit model to understand the mobilization time during hurricane evacuation." Transportation Research Part C: Emerging Technologies 32 (2013): 21-30.

Saeed, T. U., Hall, T., Baroud, H., & Volovski, M. J. (2019). Analyzing road crash frequencies with uncorrelated and correlated random-parameters count models: An empirical assessment of multilane highways. *Analytic Methods in Accident Research, 23*.

Scheiner, J., & Holz-Rau, C. (2011). A residential location approach to traffic safety: two case studies from Germany. *Accid Anal Prev, 43*(1), 307-322. doi:10.1016/j.aap.2010.08.029

Se, C., Champahom, T., Jomnonkwao, S., Chaimuang, P., & Ratanavaraha, V. (2021). Empirical comparison of the effects of urban and rural crashes on motorcyclist injury severities: A correlated random parameters ordered probit approach with heterogeneity in means. *Accid Anal Prev, 161*, 106352.

Se, C., Champahom, T., Jomnonkwao, S., Karoonsoontawon, A., & Ratanavaraha, V. (2022). Analysis

of driver-injury severity: a comparison between speeding and non-speeding driving crash accounting for temporal and unobserved effects. *Int J Inj Contr Saf Promot*, 1-14.

Se, C., Champahom, T., Jomnonkwao, S., Karoonsoontawong, A., & Ratanavaraha, V. (2021). Temporal stability of factors influencing driver-injury severities in single-vehicle crashes: A correlated random parameters with heterogeneity in means and variances approach. *Analytic Methods in Accident Research, 32*

Shafabakhsh, G. A., Famili, A., & Akbari, M. (2016). Spatial analysis of data frequency and severity of rural accidents. Transportation Letters, 1-8.

Shauna Hallmark, A. G., Theresa Litteral, Neal Hawkins, Omar Smadi, Skylar Knickerbocker. (2020). Evaluation of sequential dynamic chevron warning systems on rural two-lane curves.

*Transportation Research Record: Journal of the Transportation Research Board*.

Shyhalla, K. (2014). Alcohol involvement and other risky driver behaviors: effects on crash initiation and crash severity. *Traffic Inj Prev, 15*(4), 325-334.

Song, L., Fan, W. D., Li, Y., & Wu, P. (2021). Exploring pedestrian injury severities at pedestrian-vehicle crash hotspots with an annual upward trend: A spatiotemporal analysis with latent class random parameter approach. *J Safety Res, 76*, 184-196.

Song, L., Li, Y., Fan, W., & Wu, P. (2020). Modeling pedestrian-injury severities in pedestrian-vehicle crashes considering spatiotemporal patterns: Insights from different hierarchical Bayesian random-effects models. *Analytic Methods in Accident Research, 28*.

Songchitruksa, P., & Zeng, X. (2010). Getis–Ord Spatial Statistics to Identify Hot Spots by Using Incident Management Data. *Transportation Research Record: Journal of the Transportation Research Board, 2165*(1), 42-51.

Tay, R., Choi, J., Kattan, L., & Khan, A. (2011). A Multinomial Logit Model of Pedestrian–Vehicle Crash Severity. *International Journal of Sustainable Transportation, 5*(4), 233-249. doi:10.1080/15568318.2010.497547

Uddin, M., & Huynh, N. (2020). Injury severity analysis of truck-involved crashes under different weather conditions. *Accid Anal Prev, 141*, 105529.

Washington, S.P., Karlaftis, M.G., Mannering, F., 2010. Statistical and Econometric Methods for Transportation Data Analysis. Boca Raton, FL: CRC Press.

Wang, C., Zhang, P., Chen, F., Cheng, J., & Kim, D.-K. (2022a). Modeling Injury Severity for Nighttime and Daytime Crashes by Using Random Parameter Logit Models Accounting for Heterogeneity in Means and Variances. *Journal of Advanced Transportation, 2022*, 1-12.

Wang, C., Chen, F., Zhang, Y., Cheng, J., (2022b). Spatiotemporal instability analysis of injury severities in truck-involved and non-truck-involved crashes. *Analytic Methods in Accident Research* 34, 100214.

Wang, K., Bhowmik, T., Zhao, S., Eluru, N. and Jackson, E., 2021. Highway safety assessment and improvement through crash prediction by injury severity and vehicle damage using Multivariate Poisson-Lognormal model and Joint Negative Binomial-Generalized Ordered Probit Fractional Split model. Journal of Safety Research, 76, pp.44-55.

Wang, X., Fan, T., Chen, M., Deng, B., Wu, B., & Tremont, P. (2015). Safety modeling of urban arterials in Shanghai, China. *Accid Anal Prev, 83*, 57-66.

Wang, X., Zhou, Q., Quddus, M., Fan, T., & Fang, S. (2018). Speed, speed variation and crash relationships for urban arterials. *Accid Anal Prev, 113*, 236-243.

Wu, Q., Chen, F., Zhang, G., Liu, X. C., Wang, H., & Bogus, S. M. (2014). Mixed logit model-based driver injury severity investigations in single- and multi-vehicle crashes on rural two-lane highways. *Accid Anal Prev, 72*, 105-115.

Wang, C., Chen, F., Zhang, Y., & Cheng, J. (2022). Analysis of injury severity in rear-end crashes on an expressway involving different types of vehicles using random-parameters logit models with heterogeneity in means and variances. Transportation Letters, 1-12.

Xiao, G., Hu, Y., Li, N., & Yang, D. (2018). Spatial autocorrelation analysis of monitoring data of heavy metals in rice in China. *Food Control, 89*, 32-37.

Yamada, I., & Thill, J.-C. (2004). Comparison of planar and network K-functions in traffic accident analysis. *Journal of Transport Geography, 12*(2), 149-158.

Yan, X., He, J., Zhang, C., Liu, Z., Wang, C., & Qiao, B. (2021). Spatiotemporal instability analysis considering unobserved heterogeneity of crash-injury severities in adverse weather. *Analytic Methods in Accident Research, 32*.

Yasmin, S., & Eluru, N. (2013). Evaluating alternate discrete outcome frameworks for modeling crash injury severity. *Accid Anal Prev, 59*, 506-521.

Yuan, Renteng, et al. "Analysis of factors affecting occupant injury severity in rear-end crashes by different struck vehicle groups: A random thresholds random parameters hierarchical ordered probit model." Journal of Transportation Safety & Security (2022a): 1-22.

Yuan, Renteng, et al. "Injury severity analysis of two-vehicle crashes at unsignalized intersections using mixed logit models." International Journal of Injury Control and Safety Promotion 29.3 (2022b): 348-359.

Zeng, Q., Wang, F., Wang, Q., Pei, X., & Yuan, Q. (2022). Bayesian multivariate spatial modeling for crash frequencies by injury severity at daytime and nighttime in traffic analysis zones. Transportation Letters, 1-8

Aarts, and van Schagen. 2006. 'Driving speed and the risk of road crashes: a review', *Accid Anal Prev*, 38: 215-24.

Adanu, Agyemang, Islam, and Jones. 2021. 'A comprehensive analysis of factors that influence interstate highway crash severity in Alabama', *Journal of Transportation Safety & Security*, 14: 1552-76.

Afghari, Haque, and Washington. 2020. 'Applying a joint model of crash count and crash severity to identify road segments with high risk of fatal and serious injury crashes', *Accid Anal Prev*, 144: 105615.

Ahie, Charlton, and Starkey. 2015. 'The role of preference in speed choice', *Transportation Research Part F: Traffic Psychology and Behaviour*, 30: 66-73.

Alarifi, Abdel-Aty, and Lee. 2018. 'A Bayesian multivariate hierarchical spatial joint model for predicting crash counts by crash type at intersections and segments along corridors', *Accid Anal Prev*, 119: 263-73.

Alrejjal, Moomen, and Ksaibati. 2022. 'Evaluating the impact of traffic violations on crash injury severity on Wyoming interstates: An investigation with a random parameters model with heterogeneity in means approach', *Journal of Traffic and Transportation Engineering (English Edition)*, 9: 654-65.

Anarkooli, Hosseinpour, and Kardar. 2017. 'Investigation of factors affecting the injury severity of single-vehicle rollover crashes: A random-effects generalized ordered probit model', *Accid Anal Prev*, 106: 399-410.

Azimi, Rahimi, Asgari, and Jin. 2020. 'Severity analysis for large truck rollover crashes using a random parameter ordered logit model', *Accid Anal Prev*, 135: 105355.

Behnood, and Mannering. 2015. 'The temporal stability of factors affecting driver-injury severities in single-vehicle crashes: Some empirical evidence', *Analytic Methods in Accident Research*, 8: 7-32.

Behnood, and Mannering. 2019. 'Time-of-day variations and temporal instability of factors affecting injury severities in large-truck crashes', *Analytic Methods in Accident Research*, 23.

Bhowmik, Rahman, Yasmin, and Eluru. 2021. 'Exploring analytical, simulation-based, and hybrid model structures for multivariate crash frequency modeling', *Analytic Methods in Accident Research*, 31.

Bhowmik, Yasmin, and Eluru. 2018. 'A joint econometric approach for modeling crash counts by collision type', *Analytic Methods in Accident Research*, 19: 16-32.

Chen, Zhang, Huang, Wang, and Tarefder. 2016a. 'Examining driver injury severity outcomes in rural non-interstate roadway crashes using a hierarchical ordered logit model', *Accid Anal Prev*, 96: 79-87.

Chen, Zhang, Liu, Ci, Huang, Ma, Chen, and Guan. 2016b. 'Driver injury severity outcome analysis in rural interstate highway crashes: a two-level Bayesian logistic regression interpretation', *Accid Anal Prev*, 97: 69-78.

Dai. 2012. 'Identifying clusters and risk factors of injuries in pedestrian–vehicle crashes in a GIS environment', *Journal of Transport Geography*, 24: 206-14.

Demiroz, Onelcin, and Alver. 2015. 'Illegal road crossing behavior of pedestrians at overpass locations: Factors affecting gap acceptance, crossing times and overpass use', *Accid Anal Prev*, 80: 220-8.

Eluru, Bhat, and Hensher. 2008. 'A mixed generalized ordered response model for examining pedestrian and bicyclist injury severity level in traffic crashes', *Accid Anal Prev*, 40: 1033-54.

Fountas, Fonzone, Olowosegun, and McTigue. 2021. 'Addressing unobserved heterogeneity in the analysis of bicycle crash injuries in Scotland: A correlated random parameters ordered probit approach with heterogeneity in means', *Analytic Methods in Accident Research*, 32.

Guo, Li, Liu, and Wu. 2019. 'Modeling correlation and heterogeneity in crash rates by collision types using full bayesian random parameters multivariate Tobit model', *Accid Anal Prev*, 128: 164-74.

Hanrahan, Layde, Zhu, Guse, and Hargarten. 2009. 'The association of driver age with traffic injury severity in Wisconsin', *Traffic Inj Prev*, 10: 361-7.

Hosseinpour, Yahaya, and Sadullah. 2014. 'Exploring the effects of roadway characteristics on the frequency and severity of head-on crashes: case studies from Malaysian federal roads', *Accid Anal Prev*, 62: 209-22.

Hou, Huo, Leng, and Cheng. 2019. 'Examination of driver injury severity in freeway single-vehicle crashes using a mixed logit model with heterogeneity-in-means', *Physica A: Statistical Mechanics and its Applications*, 531.

Hoye. 2020. 'Speeding and impaired driving in fatal crashes-Results from in-depth investigations', *Traffic Inj Prev*, 21: 425-30.

Huang, Chin, and Haque. 2008. 'Severity of driver injury and vehicle damage in traffic crashes at intersections: a Bayesian hierarchical analysis', *Accid Anal Prev*, 40: 45-54.

Huang, Sun, and Tang. 2018. 'Taxi driver speeding: Who, when, where and how? A comparative study between Shanghai and New York City', *Traffic Inj Prev*, 19: 311-16.

Huang, H., Zheng, O., Wang, D., Yin, J., Wang, Z., Ding, S., ... & Shi, B. (2023). ChatGPT for Shaping the Future of Dentistry: The Potential of Multi-Modal Large Language Model. arXiv preprint arXiv:2304.03086.

Islam, and Burton. 2019. 'A comparative injury severity analysis of rural intersection crashes under different lighting conditions in Alabama', *Journal of Transportation Safety & Security*, 12: 1106-27.

Islam, and Mannering. 2021. 'The role of gender and temporal instability in driver-injury severities in crashes caused by speeds too fast for conditions', *Accid Anal Prev*, 153: 106039.

Jalayer, Shabanpour, Pour-Rouholamin, Golshani, and Zhou. 2018. 'Wrong-way driving crashes: A random-parameters ordered probit analysis of injury severity', *Accid Anal Prev*, 117: 128-35.

Le, Liu, and Lin. 2020. 'Traffic accident hotspot identification by integrating kernel density estimation and spatial autocorrelation analysis: a case study', *International Journal of Crashworthiness*, 27: 543-53.

Lee, and Li. 2014. 'Analysis of injury severity of drivers involved in single- and two-vehicle crashes on highways in Ontario', *Accid Anal Prev*, 71: 286-95.

Lemp, Kockelman, and Unnikrishnan. 2011. 'Analysis of large truck crash severity using heteroskedastic ordered probit models', *Accid Anal Prev*, 43: 370-80.

Li, and Fan. 2020. 'Mixed logit approach to modeling the severity of pedestrian-injury in pedestrian-vehicle crashes in North Carolina: Accounting for unobserved heterogeneity', *Journal of Transportation Safety & Security*, 14: 796-817.

———. 2018. 'Modelling the severity of pedestrian injury in pedestrian—vehicle crashes in North Carolina: A partial proportional odds logit model approach', *Journal of Transportation Safety & Security*, 12: 358-79.

Li, Wang, Liu, Bigham, and Ragland. 2013. 'Using Geographically Weighted Poisson Regression for county-level crash modeling in California', *Safety Science*, 58: 89-97.

Liu, and Fan. 2020. 'Analyzing injury severity of rear-end crashes involving large trucks using a mixed logit model: A case study in North Carolina', *Journal of Transportation Safety & Security*, 14: 723-36.

Lym, and Chen. 2021. 'Influence of built environment on the severity of vehicle crashes caused by distracted driving: A multi-state comparison', *Accid Anal Prev*, 150: 105920.

Mannering, Shankar, and Bhat. 2016. 'Unobserved heterogeneity and the statistical analysis of highway accident data', *Analytic Methods in Accident Research*, 11: 1-16.

Mohamad, Abdullah, and Mohamad. 2019. 'Are sociodemographic characteristics and attitude good predictors of speeding behavior among drivers on Malaysia federal roads?', *Traffic Inj Prev*, 20: 478-83.

NHTSA. 2020. 'United States Department of Transportation'. https://www.nhtsa.gov/risky-driving/speeding.

Osman, Mishra, and Paleti. 2018. 'Injury severity analysis of commercially-licensed drivers in single-vehicle crashes: Accounting for unobserved heterogeneity and age group differences', *Accid Anal Prev*, 118: 289-300.

Park, Yang, Park, and Kim. 2020. 'Multiple membership multilevel model to estimate intersection crashes', *Accid Anal Prev*, 144: 105589.

Peng, Wang, Peng, Huang, Tian, and Jia. 2018. 'Investigation on the injuries of drivers and copilots in rear-end crashes between trucks based on real world accident data in China', *Future Generation Computer Systems*, 86: 1251-58.

Plug, Xia, and Caulfield. 2011. 'Spatial and temporal visualisation techniques for crash analysis', *Accid Anal Prev*, 43: 1937-46.

Pulugurtha, Krishnakumar, and Nambisan. 2007. 'New methods to identify and rank high pedestrian crash zones: an illustration', *Accid Anal Prev*, 39: 800-11.

Qi, Srinivasan, Teng, and Baker. 2013. 'Analysis of the frequency and severity of rear-end crashes in work zones', *Traffic Inj Prev*, 14: 61-72.

Rahman, Zafri, Akter, and Pervaz. 2021. 'Identification of factors influencing severity of motorcycle crashes in Dhaka, Bangladesh using binary logistic regression model', *Int J Inj Contr Saf Promot*, 28: 141-52.

Rezapour, and Ksaibati. 2022. 'Contributory factors to the severity of single-vehicle rollover crashes on a mountainous area, generalized additive model', *Int J Inj Contr Saf Promot*, 29: 281-88.

Rezapour, Moomen, and Ksaibati. 2019. 'Ordered logistic models of influencing factors on crash injury severity of single and multiple-vehicle downgrade crashes: A case study in Wyoming', *J Safety Res*, 68: 107-18.

Sadri, Ukkusuri, and Murray-Tuite. 2013. 'A random parameter ordered probit model to understand the mobilization time during hurricane evacuation', *Transportation Research Part C: Emerging Technologies*, 32: 21-30.

Saeed, Hall, Baroud, and Volovski. 2019. 'Analyzing road crash frequencies with uncorrelated and correlated random-parameters count models: An empirical assessment of multilane highways', *Analytic Methods in Accident Research*, 23.

Scheiner, and Holz-Rau. 2011. 'A residential location approach to traffic safety: two case studies from Germany', *Accid Anal Prev*, 43: 307-22.

Se, Champahom, Jomnonkwao, Chaimuang, and Ratanavaraha. 2021a. 'Empirical comparison of the effects of urban and rural crashes on motorcyclist injury severities: A correlated random parameters ordered probit approach with heterogeneity in means', *Accid Anal Prev*, 161: 106352.

Se, Champahom, Jomnonkwao, Karoonsoontawon, and Ratanavaraha. 2022. 'Analysis of driver-injury severity: a comparison between speeding and non-speeding driving crash accounting for temporal and unobserved effects', *Int J Inj Contr Saf Promot*, 29: 475-88.

Se, Champahom, Jomnonkwao, Karoonsoontawong, and Ratanavaraha. 2021b. 'Temporal stability of factors influencing driver-injury severities in single-vehicle crashes: A correlated random parameters with heterogeneity in means and variances approach', *Analytic Methods in Accident Research*, 32.

Shyhalla. 2014. 'Alcohol involvement and other risky driver behaviors: effects on crash initiation and crash severity', *Traffic Inj Prev*, 15: 325-34.

Song, Fan, Li, and Wu. 2021a. 'Exploring pedestrian injury severities at pedestrian-vehicle crash hotspots with an annual upward trend: A spatiotemporal analysis with latent class random parameter approach', *J Safety Res*, 76: 184-96.

Song, Li, Fan, and Liu. 2021b. 'Mixed logit approach to analyzing pedestrian injury severity in pedestrian-vehicle crashes in North Carolina: Considering time-of-day and day-of-week',

*Traffic Inj Prev*, 22: 524-29.

Song, Li, Fan, and Wu. 2020. 'Modeling pedestrian-injury severities in pedestrian-vehicle crashes considering spatiotemporal patterns: Insights from different hierarchical Bayesian random-effects models', *Analytic Methods in Accident Research*, 28.

Songchitruksa, and Zeng. 2010. 'Getis–Ord Spatial Statistics to Identify Hot Spots by Using Incident Management Data', *Transportation Research Record: Journal of the Transportation Research Board*, 2165: 42-51.

Tay, Choi, Kattan, and Khan. 2011. 'A Multinomial Logit Model of Pedestrian–Vehicle Crash Severity', *International Journal of Sustainable Transportation*, 5: 233-49.

Uddin, and Huynh. 2020. 'Injury severity analysis of truck-involved crashes under different weather conditions', *Accid Anal Prev*, 141: 105529.

Wang, Bhowmik, Zhao, Eluru, and Jackson. 2021. 'Highway safety assessment and improvement through crash prediction by injury severity and vehicle damage using Multivariate Poisson-Lognormal model and Joint Negative Binomial-Generalized Ordered Probit Fractional Split model', *J Safety Res*, 76: 44-55.

Wang, Chen, Zhang, Wang, Yu, and Cheng. 2022a. 'Temporal stability of factors affecting injury severity in rear-end and non-rear-end crashes: A random parameter approach with heterogeneity in means and variances', *Analytic Methods in Accident Research*, 35.

Wang, Fan, Chen, Deng, Wu, and Tremont. 2015. 'Safety modeling of urban arterials in Shanghai, China', *Accid Anal Prev*, 83: 57-66.

Wang, Zhang, Chen, Cheng, and Kim. 2022b. 'Modeling Injury Severity for Nighttime and Daytime Crashes by Using Random Parameter Logit Models Accounting for Heterogeneity in Means and Variances', *Journal of Advanced Transportation*, 2022: 1-12.

Wang, Zhou, Quddus, Fan, and Fang. 2018. 'Speed, speed variation and crash relationships for urban arterials', *Accid Anal Prev*, 113: 236-43.

Wu, Chen, Zhang, Liu, Wang, and Bogus. 2014. 'Mixed logit model-based driver injury severity investigations in single- and multi-vehicle crashes on rural two-lane highways', *Accid Anal Prev*, 72: 105-15.

Xiao, Hu, Li, and Yang. 2018. 'Spatial autocorrelation analysis of monitoring data of heavy metals in rice in China', *Food Control*, 89: 32-37.

Yamada, and Thill. 2004. 'Comparison of planar and network K-functions in traffic accident analysis', *Journal of Transport Geography*, 12: 149-58.

Yan, He, Wu, Zhang, Liu, and Wang. 2021a. 'Weekly variations and temporal instability of determinants influencing alcohol-impaired driving crashes: A random thresholds random parameters hierarchical ordered probit model', *Analytic Methods in Accident Research*, 32.

Yan, He, Zhang, Liu, Wang, and Qiao. 2021b. 'Spatiotemporal instability analysis considering unobserved heterogeneity of crash-injury severities in adverse weather', *Analytic Methods in Accident Research*, 32.

Yasmin, and Eluru. 2013. 'Evaluating alternate discrete outcome frameworks for modeling crash injury severity', *Accid Anal Prev*, 59: 506-21.

Yuan, Gan, Peng, and Xiang. 2022a. 'Injury severity analysis of two-vehicle crashes at unsignalized intersections using mixed logit models', *Int J Inj Contr Saf Promot*, 29: 348-59.

Yuan, Gu, Peng, and Xiang. 2022b. 'Analysis of factors affecting occupant injury severity in rear-end crashes by different struck vehicle groups: A random thresholds random parameters hierarchical ordered probit model', *Journal of Transportation Safety & Security*: 1-22.

Zeng, Wang, Wang, Pei, and Yuan. 2022. 'Bayesian multivariate spatial modeling for crash frequencies by injury severity at daytime and nighttime in traffic analysis zones', *Transportation Letters*: 1-8.

Zhang, and Hassan. 2019a. 'Crash severity analysis of nighttime and daytime highway work zone crashes', *PLoS One*, 14: e0221128.

Zhang, and Hassan. 2019b. 'Identifying the Factors Contributing to Injury Severity in Work Zone Rear-End Crashes', *Journal of Advanced Transportation*, 2019: 1-9.

Zheng, O., Abdel-Aty, M., Wang, D., Wang, Z., & Ding, S. (2023). ChatGPT is on the horizon: Could a large language model be all we need for Intelligent Transportation?. arXiv preprint arXiv:2303.05382.